# Critical role of terminating layer in formation of 2DEG state at the LaInO$_3$/BaSnO$_3$ interface

Seonghyeon Kim, Mikk Lippmaa, Jaehyeok Lee, Hyeongmin Cho, Juhan Kim, Bongju Kim[*] and Kookrin Char[*]

S. Kim, J. Lee, J. Kim, Prof. K. Char[*]

Institute of Applied Physics, Department of Physics and Astronomy, Seoul National University; Seoul 08826, Republic of Korea

[*]E-mail: kchar@snu.ac.kr

Prof. M. Lippmaa

Institute for Solid State Physics, University of Tokyo, Kashiwa 277-8581, Japan

Dr. B. Kim[*],

Center for Correlated Electron Systems (CCES), Institute for Basic Science (IBS), Seoul 08826, Republic of Korea

Institute of Applied Physics, Department of Physics and Astronomy, Seoul National University; Seoul 08826, Republic of Korea

[*]E-mail: bongju@snu.ac.kr

**Abstract:** Based on the interface polarization model, the two-dimensional electron gas (2DEG) at LaInO$_3$(LIO)/BaSnO$_3$(BSO) interfaces is understood to originate from a polarization discontinuity at the interface and the conduction band offset between LIO and BSO. In this scenario, the direction of polarization at the interface is determined by whether the first atomic LIO layer at the interface is LaO$^+$ or InO$_2^-$. We investigate the role of the terminating layer at the LIO/BSO interface in creating the 2DEG. Based on conductance measurements of our in-situ grown LIO/BSO heterostructures, we report in this work that the 2DEG only forms when the BSO surface is terminated with a SnO$_2$ layer. We controlled the terminating layer by additional SnO$_2$ deposition on the BSO surface. We show that the as-grown BSO surface has a mixed terminating layer of BaO and SnO$_2$ while the BSO surfaces prepared with additional SnO$_2$ deposition are terminated mainly with the SnO$_2$ layer. The terminating layer was confirmed by coaxial impact collision ion scattering spectroscopy (CAICISS). Our finding is consistent with the interface polarization model for 2DEG formation at LIO/BSO interfaces, in which the direction of the interfacial polarization in LIO is determined by the terminating layer of the BSO surface.

**Introduction**

The two-dimensional electron gas (2DEG) where electrons are confined in a narrow quantum well has been a very fertile ground for discovering novel phenomena as well as for use in semiconductor applications such as high electron mobility transistors (HEMTs).[1,2] The quantum well in conventional 2DEGs (AlGaAs/GaAs, AlGaN/GaN, and MgZnO/ZnO) is created by the conduction band offset between the two materials.[1–4] Two different approaches have been studied for conventional semiconductor 2DEGs: the modulation doping approach is used for GaAs while polarization-induced carrier accumulation occurs in GaN and ZnO.[1,2,5,6] On the other hand, 2DEG-like behavior was also found at LaAlO$_3$(LAO)/SrTiO$_3$(STO) interfaces in 2004.[7] Many properties of the LAO/STO 2DEGs were found to be different from the conventional semiconductor 2DEGs like AlGaAs/GaAs, AlGaN/GaN and MgZnO/ZnO, as was the formation mechanism. The 'polar catastrophe', and the resulting charge transfer mechanism through the interface, has been considered as the primary mechanism for 2DEG between LAO and STO.[8] Recently, sheet conductance enhancement by four order of magnitudes was reported at the LIO/BSO interface.[9,10] The LIO/BSO 2DEG is described by a model in which polar discontinuity occurs only near the interface ('interface polarization' model) due to the inversion symmetry breaking at the interface.[9,11]

Conventional 2DEG - AlGaAs/GaAs

From the 1980s, the AlGaAs/GaAs 2DEG has been widely studied for HEMT applications.[1,2] The high mobility of electrons in AlGaAs/GaAs 2DEGs is achieved by reducing impurity scattering through modulation doping. Since GaAs does not suffer from unintentional doping, the electron mobility in GaAs can be increased by modulation doping the larger bandgap AlGaAs with a Si dopant. In the AlGaAs/GaAs system, a quantum well is formed at the interface due to the conduction band offset between the two materials.[2] GaAs has the zinc blende structure in the $F\bar{4}3m$ space group and it is thus a non-centrosymmetric crystal. Despite the non-centrosymmetric crystal structure, the spontaneous polarization of GaAs and AlGaAs is almost zero.[12,13] The piezoelectric polarization is negligible because the lattice constant of AlGaAs is nearly the same as that of GaAs.[12,13] Since there is no polarization in GaAs and AlGaAs, no polarization discontinuity exits at the interface, and a quantum well is formed only by the conduction band offset. Therefore, in the AlGaAs/GaAs system, the effect of crystalline direction or terminating layer, which may determine the direction of the polarization, can be ignored.

Conventional 2DEG – Polarization induced 2DEG

The AlGaN/GaN and MgZnO/ZnO interfaces form 2DEGs due to polarization-induced doping.[5,6] Both GaN and ZnO have the wurtzite structure belonging to the $P6_3mc$ space group. The crystals consist of hexagonally arranged Ga-N binary layers in an alternating stacking sequence, which is non-centrosymmetric. If the crystal direction is [0001], the binary layers are terminated with Ga and it becomes Ga-face. If it is the opposite direction, it becomes N-face. Because of their non-centrosymmetricity, GaN has a spontaneous polarization of 2.90 µC cm$^{-2}$ and Al$_{0.32}$Ga$_{0.68}$N has a spontaneous polarization ($P_{sp}$) of 4.56 µC cm$^{-2}$.[5,12] The direction of the spontaneous polarization is determined by the crystal orientation due to the inversion symmetry breaking. When the wurtzite lattice is viewed along the [0001] or [000$\bar{1}$] crystallographic directions, the dipoles between the two constituent atoms are all aligned in one direction, determining the polarization direction. The Ga face [0001] presents spontaneous polarization toward the GaN from AlGaN, and the N face [000$\bar{1}$] presents spontaneous polarization in opposite direction.[5] In addition to the spontaneous polarization, there exists the piezoelectric polarization ($P_{pe}$) effect due to the lattice mismatch between GaN and AlGaN in a AlGaN/GaN heterostructure.[5] The direction of the piezoelectric polarization is determined by the direction of the spontaneous polarization and the type of strain in each layer. The $P_{sp}$ and $P_{pe}$ of GaAs, AlGaAs, GaN, AlGaN, ZnO, and MgZnO can be found in Table S1 in Supporting Information, where the ratio of Al and Mg alloying is 32%, respectively. As shown in Table S1, the total polarization in AlGaN is $P_{tot} = P_{sp} + P_{pe}$ = 4.56 µC cm$^{-2}$ + 1.17 µC cm$^{-2}$ = 5.73 µC cm$^{-2}$ at its maximum when AlGaN is under tensile strain on relaxed GaN, and 4.56 µC cm$^{-2}$ at its minimum when the AlGaN layer is fully relaxed on GaN.[5,12,14] Therefore, the polarization discontinuity ($\Delta P$) at the interface of AlGaN/GaN is $\Delta P = P_{AlGaN} - P_{GaN}$ = 2.83 µC cm$^{-2}$ at its maximum and $\Delta P$ = 1.66 µC cm$^{-2}$ at its minimum. In the case of the GaN/AlGaN structure, the GaN layer can be subjected to compressive strain on relaxed AlGaN. In this case, $\Delta P$ becomes -2.70 $\mu$C cm$^{-2}$. In addition to the polarization discontinuity at the interface, the formation of a 2DEG at the interface is aided by the conduction band offset, which is 1.1 eV for the Al$_{0.32}$GaN$_{0.68}$/GaN interface. Unlike an AlGaAs/GaAs 2DEG, intentional doping in the wider bandgap AlGaN is not required due to the existence of intrinsic deep donors. The N vacancies as well as the O and Si impurities in GaN were considered as native defects generating intrinsic donors and these donors become deep donors when the Al ratio in AlGaAs is above 0.4.[15-17] In the case of ZnO, Zn and O are playing the same role as Ga and N in GaN.[8,18]

Polar catastrophe model – LAO/STO

The very high sheet carrier density found at LAO/STO interfaces, larger than $3 \times 10^{14}$ cm$^{-2}$, is much higher than those in the conventional semiconductor 2DEGs.[19] The origin of such high carrier density has not been fully understood yet, but several origins have been proposed, such as oxygen vacancies, interfacial mixing and the polar catastrophe.[8,20-22] Among these, the polar catastrophe model is widely accepted. According to the polar catastrophe model, the internal potential diverges as the thickness of the LAO layer increases but the divergence catastrophe can be avoided if half an electron per unit cell ($3 \times 10^{14}$ cm$^{-2}$) charge transfer across the interface occurs and it compensate electrostatic potential at the (LaO)$^+$/(TiO$_2$)$^0$ interface.[8] This is why the polar catastrophe model is called a 'charge transfer' model.[8] In this model, the (AlO$_2$)$^-$/(SrO)$^0$ interface is expected to form a two-dimensional hole gas (2DHG), which has been reported to be experimentally observed.[23] The termination of the interface, either (AlO$_2$)$^-$/(SrO)$^0$ or (LaO)$^+$/(TiO$_2$)$^0$ thus determines whether a 2DEG or a 2DHG forms at the interface in this model.

Interface polarization - LIO/BSO

BaSnO$_3$ (BSO) is a cubic perovskite semiconductor with a wide band gap (3.1 eV).[24,25] LaInO$_3$ (LIO) is an orthorhombic perovskite with a large band gap of 5.0 eV and its pseudocubic lattice constant is 4.117 Å which is almost matched to the BSO cubic lattice constant of 4.116 Å.[26,27] The space group of BSO is $Pm\bar{3}m$ which is centrosymmetric. The orthorhombic LIO is in the space group of *Pnma* which is also centrosymmetric. Therefore, neither LIO nor BSO can have spontaneous polarization in their bulk form. However, within a few unit cells near the interface where the translation symmetry is broken, a polar state can exist in the growth direction in LIO. Therefore, an "interfacial polarization" model has been proposed to explain the 2DEG formation at LIO/BSO interfaces.[9] Experimentally, the LIO/BSO conductance reaches a maximum value when the LIO thickness is four unit cells and starts to decrease for thicker layers.[10,11] It has been shown that this experimental result can be explained by the 'interface polarization' model through a self-consistent Poisson-Schrödinger (P-S) calculation.[11,12]

As shown in **Figure 1**a, the BSO lattice consists of a sequence of nonpolar layers of $(BaO)^0$ and $(SnO_2)^0$. Since the $(BaO)^0$ and $(SnO_2)^0$ layers in BSO are neutral, spontaneous polarization is not possible. On the other hand, the LIO has a sequence of polar layers of $(LaO)^+$ and $(InO_2)^-$. The $(LaO)^+/(InO_2)^-$ and $(InO_2)^-/(LaO)^+$ dipoles cancel in the overall bulk due to its centrosymmetry. However, the dipoles near the interface do not cancel out and can create polarization. At the interface, there are two possible layer sequences: $(InO_2)^-/(LaO)^+/(SnO_2)^0$ and $(LaO)^+/(InO_2)^-/(BaO)^0$. In the former case, the polarization points towards BSO and in the latter case the polarization is towards LIO. Moreover, it has been reported that when orthorhombic LIO is grown on cubic BSO, the strained LIO layer near the interface exhibits a suppression of octahedral tilting.[11,28] Figure 1b shows a self-consistent Poisson-Schrödinger (P-S) calculation result for the band bending near the LIO/BSO interface for opposite polarization states. We used the P-S calculator by Snider.[29] The materials parameters of BSO and LIO that were used in the P-S calculation are shown in Table S2 in Supporting Information. In the case of $SnO_2$ termination $(InO_2)^-/(LaO)^+/(SnO_2)^0$, a 2DEG was formed at the interface, but for BaO termination $(LaO)^+/(InO_2)^-/(BaO)^0$, no 2DEG can be formed, nor can a 2DHG form due to the very small valence band offset. Since LIO/BSO 2DEG formation is determined by the LIO polarization direction at the interface, the 2DEG formation will be sensitive to the terminating layer of the BSO crystal. To examine this, we studied the LIO/BSO conductance by controlling the termination of the BSO surface via in-situ additional $SnO_2$ depositions.

**Experiment**

We prepared the samples by pulsed laser deposition with a KrF excimer laser with a 248 nm wavelength. The fluence on the target was in the range of 1.4~1.5 J/cm$^2$ and the deposition was performed at 750 °C and 100 mTorr oxygen pressure. The samples were cooled down in 600 Torr of oxygen after deposition to room temperature. The targets were supplied by Toshima Manufacturing Co.. To make electrical contacts, indium contacts were pressed on the 4 corners of a sample in the Van der Pauw geometry. The electrical measurement was carried out with Keithley 4200 SCS. All electrical measurement was taken at room temperature.

**Results and Discussion**

Growth rate measurement of $SnO_2$ with XRR

**Figure 2** shows a $SnO_2$ growth rate measurement by X-ray reflectometry (XRR). The $SnO_2$ film was grown in the (101) direction of the rutile structure ($a = b = 4.74$ Å, $c = 3.18$ Å) on a ($10\bar{1}2$) $Al_2O_3$ substrate.[30] As shown Figure 2, the thickness of a $SnO_2$ film grown with 525 laser pulses was 204 Å, determined by analyzing the positions of maxima of Kiessig fringes. The growth rate of $SnO_2$ was thus 0.384 Å per 1 pulse of laser deposition. To evaluate the growth direction and quality of $SnO_2$, we measured an X-ray diffraction pattern of a 150 nm thick $SnO_2$ film on ($10\bar{1}2$) $Al_2O_3$ as shown in the inset of Figure 2. In order to convert the measured $SnO_2$ growth rate to the layer deposition rate of perovskite-type BSO, we considered the unit cell volume difference of each crystal structure. Two Sn atoms are contained in each unit cell volume (71.51 Å$^3$) of the rutile structure while a single Sn atom is contained in a half unit cell of perovskite BSO (34.87 Å$^3$). By taking into account the packing density difference between the rutile and perovskite $SnO_2$, the growth rate of the $SnO_2$ layer for the perovskite structure was adjusted to be 0.379 Å per 1 pulse of laser deposition. For example, to obtain a single monolayer of $SnO_2$ with the perovskite structure, a 6-pulse laser deposition of $SnO_2$ (2.27 Å) is needed. If an as-grown BSO surface has a ratio of 50:50 of BaO and $SnO_2$, its surface can be changed to mainly $SnO_2$-terminated BSO with 1.03 Å of $SnO_2$ or 3-pulse laser deposition (1.14 Å), assuming all the $SnO_2$ ends up on top of the BaO layer.

### 2DEG conductance as a function of SnO$_2$ thickness (on MgO)

To examine the 2DEG conductance as a function of SnO$_2$ thickness, we prepared differently terminated BSO films by an additional SnO$_2$ deposition on as-grown BSO. As shown **Figure 3**a, we deposited the in-situ LIO/BSO samples on top of an 80 nm thick BaHfO$_3$ (BHO) buffer layer on MgO substrates. The BHO buffer layer was used for strain relaxation since its cubic lattice constant 4.171 Å is between that of MgO (4.212 Å) and that of BSO (4.116 Å).[24,31] After depositing a 200 nm thick BSO layer, we deposited a few additional pulses of SnO$_2$ on the BSO surface for termination control, followed by 10 nm thick LIO deposition. As shown in Figure 3b, the heterostructure grown without additional SnO$_2$ deposition at the interface was basically insulating, with a sheet conductance of $<1\times10^{-11}$ $\Omega^{-1}$. The highest sheet conductance of $2.56 \times10^{-6}$ $\Omega^{-1}$ was achieved with a 3-pulse (1.1 Å) SnO$_2$ termination control deposition at the interface. The sheet conductance slowly decreased to $3.64 \times10^{-7}$ $\Omega^{-1}$ for a 5-pulse (1.9 Å) SnO$_2$ deposition and then rapidly dropped to $7.79 \times10^{-10}$ $\Omega^{-1}$ at 6 pulses of SnO$_2$. The highest sheet conductance measured here is about 2 times higher than what has been reported in the past for samples grown on MgO in an ex-situ process.[10] The dramatic effect of the SnO$_2$ deposition at the interface on the sheet conductance shows that the termination control of the BSO surface is crucial for obtaining a high-mobility 2DEG at a LIO/BSO interface. Considering that the peak point is at the 3 pulse deposition (1.1 Å) of SnO$_2$, the optimal SnO$_2$ layer thickness is 1.1 Å, which suggests that the Ba:Sn ratio of the thick as-grown BSO film is close to 50:50. In such case the surface of the BSO film requires about 1/4 unit cell (1.029 Å) of SnO$_2$ to complete the SnO$_2$ termination layer. This is consistent with an ab initio calculation which reported that both SnO$_2$ and BaO terminations on the BSO surface are energetically equally stable.[32]

### 2DEG conductance as a function of SnO$_2$ thickness (on SrTiO$_3$)

We also investigated the conductance of LIO/BSO 2DEGs on SrTiO$_3$ (STO) substrates. Figure 3c shows a schematic diagram of these samples. The 10 nm LIO was deposited on 200 nm thick nondoped BSO with a few pulses of SnO$_2$ deposition at the interface. The 2DEG conductance as a function of SnO$_2$ deposition thickness is shown in Figure 3d. The maximum 2DEG conductance was achieved at 1.1 Å of $SnO_2$ and decreases thereafter. This trend is similar to the samples made on MgO substrates shown in Figure 3b. This result shows the choice of substrate has no significant effect on the surface composition of as-grown BSO films. However, the highest sheet conductance of $4.47 \times 10^{-7}$ $\Omega^{-1}$ on the STO substrate is about 6 times lower than what was obtained on MgO substrates, as shown in Figure 3b. This difference can be attributed to the different deep acceptor density of BSO films grown on different substrates.[10] The deep acceptor density in BSO grown on MgO was $4 \times 10^{19}$ $cm^{-3}$ and $6 \times 10^{19}$ $cm^{-3}$ on STO.[11,12,33] In our earlier work, we reported that the LIO/BSO(undoped) heterostructures on STO, made by ex-situ processes, were insulating with the sheet conductance lower than $10^{-10}$ $\Omega^{-1}$.[9] In contrast, here we report a sheet conductance of LIO/BSO(undoped) on STO as high as $4.47 \times 10^{-7}$ $\Omega^{-1}$ when made by the in-situ $SnO_2$ termination control process.

Coaxial Impact Collision Ion Scattering (CAICISS)

To analyze the composition of the BSO surface, we used coaxial impact collision ion scattering (CAICISS), (Shimadzu, TALIS-9700) analyzer. As shown **Figure 4**a, an incident $Ne^+$ ion beam was aligned with the [111] direction of BSO. Due to the incident beam direction alignment, the second layer from the top will be hidden from the ion beam by the shadow cone in a perovskite structure except for the exposed edges. We measured the time-of-flight (TOF) of backscattered $Ne^+$, which was only affected by the atomic mass of the scattering cations of the BSO surface.[34] Figure 4b shows the TOF spectra of three different samples. In the case of the as-grown BSO, backscattering peaks from both Ba and Sn are equally large. For the sample with an additional 1.1 Å $SnO_2$ deposition, the Ba peak intensity decreased significantly, showing that the additional $SnO_2$ deposition does indeed change the terminating layer composition of the BSO surface. However, the Ba peak did not completely disappear even after 1.9 Å thick $SnO_2$ deposition. As shown in the AFM images in the supplemental material (Figure S1, Supporting Information), this is likely due to the fact that the sample surface isn't atomically smooth, exposing Ba atoms on the edges of the small islands to the $Ne^+$ ion beam.

STEM images of LIO/BSO with different $SnO_2$ thickness (1.1 Å and 4.2 Å)

**Figure 5**a,b show the high angle annular dark field scanning transmission electron microscope (HAADF-STEM) cross section images of an LIO film grown on a BSO film terminated with an optimal 3-pulse deposition of a 1.1 Å $SnO_2$ layer at the interface. Figure 5d,e show the effect of depositing an additional 4.2 Å (11 pulses) $SnO_2$ layer on BSO. Figure 5c shows a schematic diagram of the sample shown in Figure 5a,b and Figure 5f shows a schematic diagram of the sample corresponding to Figs. 5d,e. In both cases, LIO was epitaxially grown on BSO. Since the contrast between Ba and La and between Sn and In is very small, the LIO/BSO interface in most previous reports was not easy to see.[9-11,35] However, the interface is more clearly visible at lower magnification, as can be seen by comparing images in Figure 5a,d. The higher magnification images in Figure 5b,e show clearly that the growth of subsequent LIO remains epitaxial regardless of the additional $SnO_2$ deposition thickness. However, there exist areas with more visible darker contrast in Figure 5e near the interface which suggests that defect-related strain exists near the interface. For the thicker $SnO_2$ layer, the off-stoichiometry can be accommodated by creating defects such as $Sn_{Ba}$ or $V_{Ba}^{''}$, which are known to form during the growth of BSO in Sn-rich conditions of molecular beam epitaxy.[36] Although the film maintains epitaxial growth at the interface, the sheet conductance with 4.2 Å $SnO_2$ deposition was just $1.52 \times 10^{-9}$ $\Omega^{-1}$, which is 3 orders of magnitude lower than the highest conductance of $2.56 \times 10^{-6}$ $\Omega^{-1}$, obtained with the optimal $SnO_2$ coverage shown in Fig. 3(b).

Comparison with previous ex-situ samples

In our previous work [9-11,35] on LIO/BSO interfaces all samples were made in an ex-situ process unlike the in-situ deposition described here for all the layers involved in forming the interface. In the ex-situ process, the BSO surface is exposed to air and then heated back to high temperature twice for the exchange of stencil masks. It is very likely that during the air exposure and the reheating process the film surface composition may change. The very hygroscopic BaO easily forms $Ba(OH)_2$ in humid air, which will evaporate in the subsequent heating process due to the low boiling temperature of 780 °C.

**Conclusion**

In summary, we have shown that the use of $SnO_2$-terminated BSO is crucial for obtaining a conducting interface at a LIO/BSO interface. It is very likely that the polarization direction at the LIO/BSO interface is sensitively affected by the termination of the BSO. We have shown that the BSO termination could be controlled by additional $SnO_2$ deposition of just a few PLD laser pulses. The electrical conductance increased by more than 4 orders of magnitude with optimal $SnO_2$ termination. We confirmed the composition of the BSO surfaces by CAICISS. Since the optimal sheet conductance was found when 1/4 unit cell (1.1 Å) of $SnO_2$ was added at the interface, the as-grown BSO surface was found to have nearly equal amounts of BaO and $SnO_2$ as the terminating layer after PLD deposition.

**Acknowledgement**

a

SnO$_2$ termination

BaO termination

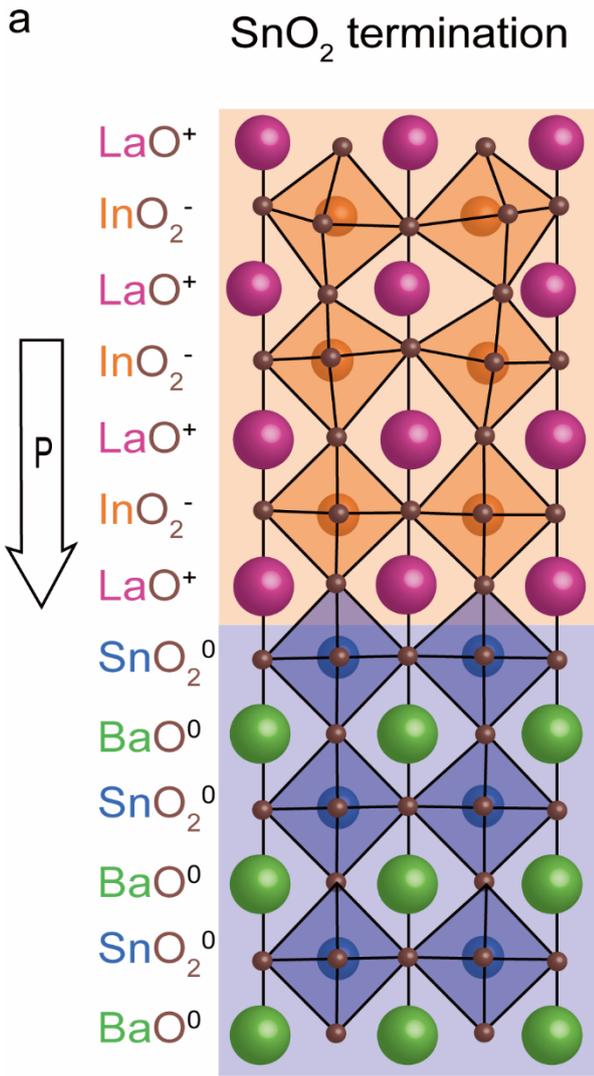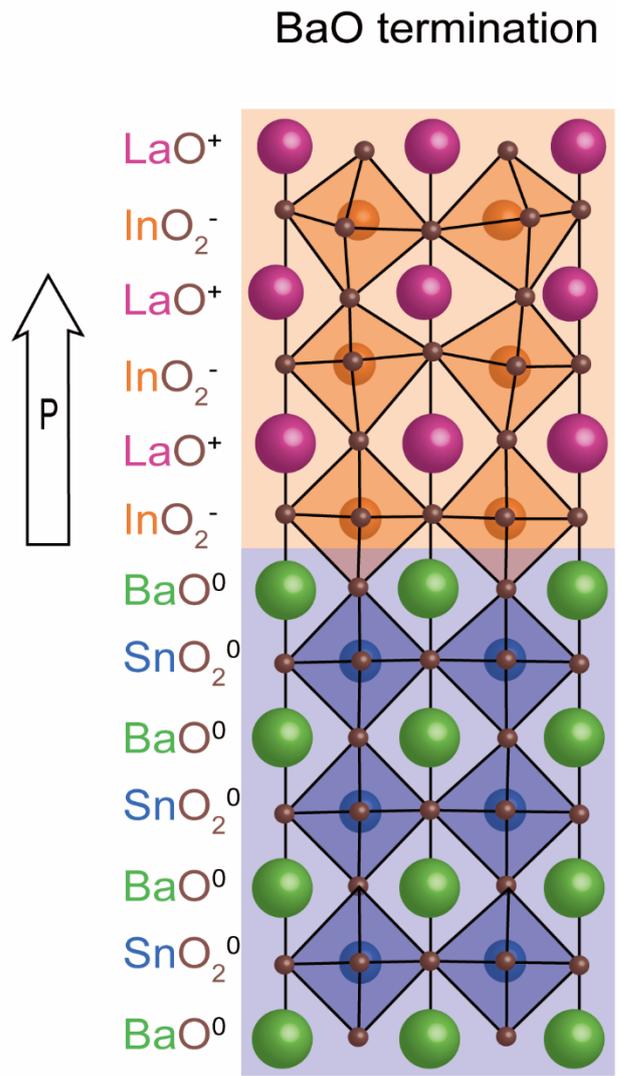

b

SnO$_2$ termination

BaO termination

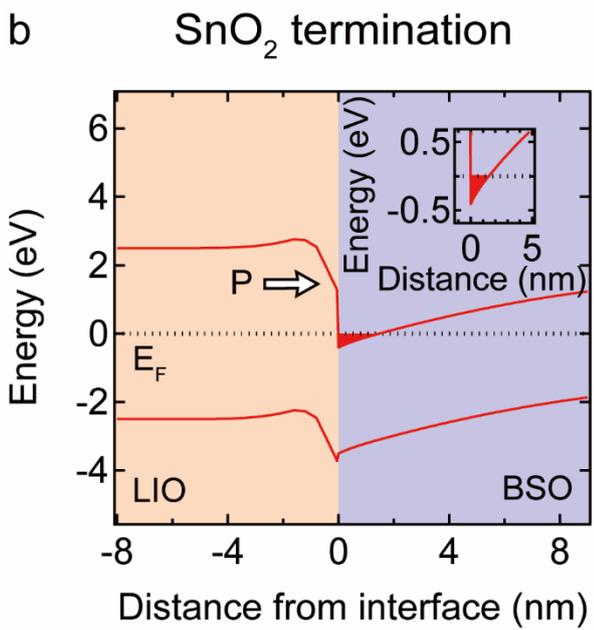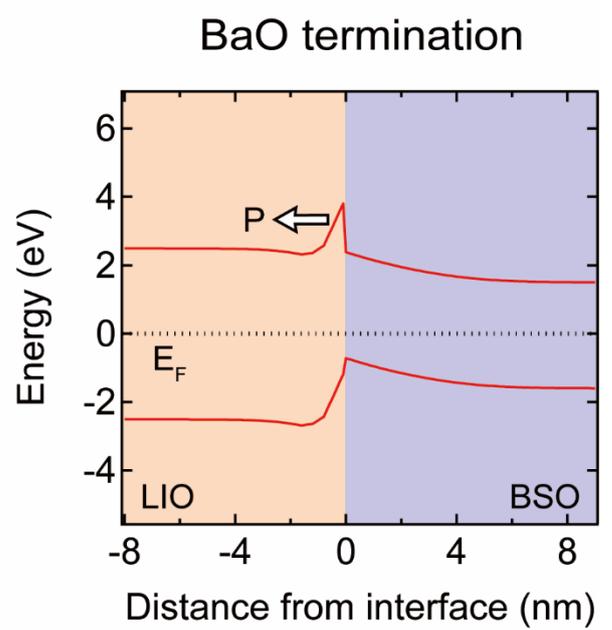

**Figure 1.** a) Schematic illustration of the $(LaO)^+/(SnO_2)^0$ interface ($SnO_2$ termination) and $(InO_2)^-/(BaO)^0$ interface (BaO termination). At the $(LaO)^+/(SnO_2)^0$ interface, direction of the polarization is toward to $BaSnO_3$. At the $(InO_2)^-/(BaO)^0$ interface, direction of the polarization is the reverse of the previous one. b) Calculated band diagram at the $SnO_2$ termination and BaO termiation interface using a self-consistent P-S calculation. In addition to the conduction band offset between the two materials, energy level is lowered due to polarization at the interface. In case of $SnO_2$ termination, 2DEG was formed at the interface, but in case of BaO termination neither 2DEG nor 2DHG can be formed.

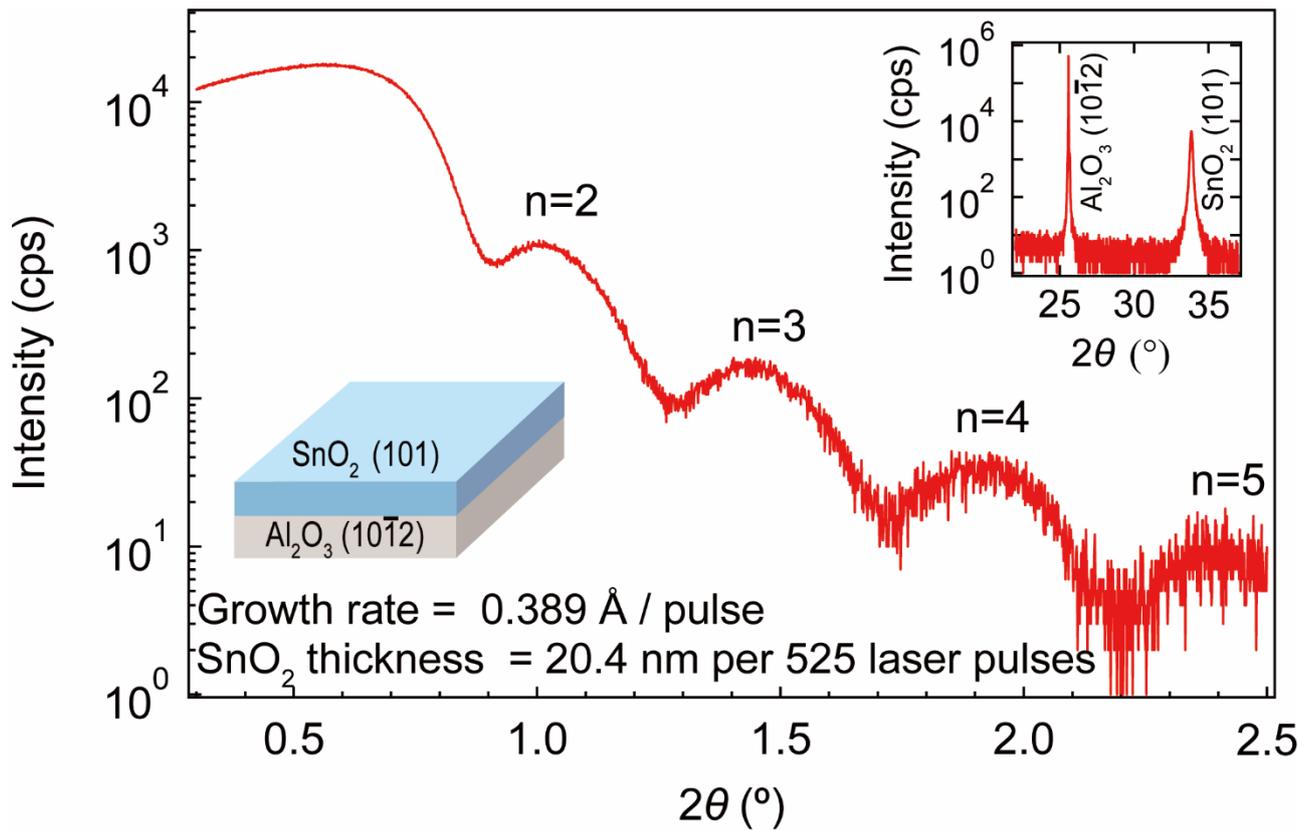

**Figure 2.** SnO$_2$ X-ray spectrum and growth rate measurement using X-ray reflectivity (XRR). SnO$_2$ is grown by 525 laser pulses with the laser energy fluence of 1.4 J/cm$^2$ in 100 mTorr O$_2$ atmosphere at 750 °C. The results of X-ray reflectivity measurement of SnO$_2$ sample. Inset presents X-ray diffraction spectrum of 150 nm SnO$_2$ on (10$\bar{1}$2) Al$_2$O$_3$ substrate. The growth direction of SnO$_2$ is (101) direction.

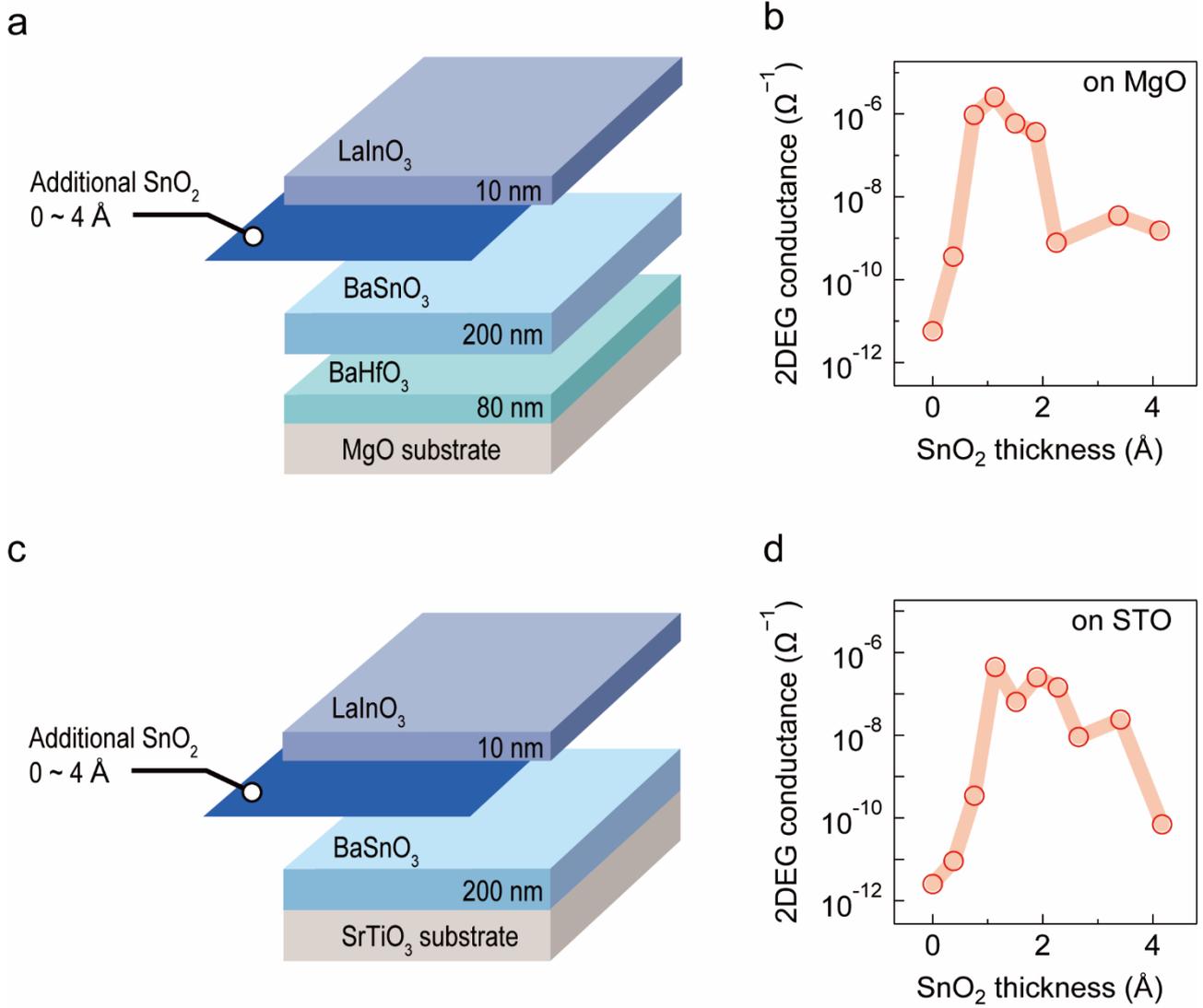

**Figure 3.** LIO/BSO 2DEG conductance as a function of additional $SnO_2$. a) Schematic illustration of in-situ LIO/BSO 2DEG on MgO with additional $SnO_2$ deposition on BSO layer. b) The sheet conductance of LIO/BSO on MgO substrate as a function of additional $SnO_2$ thickness. The 2DEG conductance increases until 1.1 Å of $SnO_2$ and the conductance decreases as thickness of $SnO_2$ increases beyond 1.1 Å. c) Schematic illustration of in-situ LIO/BSO 2DEG on STO with $SnO_2$ dusting on BSO layer. d) The sheet conductance of LIO/BSO on STO substrate as a function of additional $SnO_2$ thickness. This trend was similar to the samples made on MgO substrates in (b).

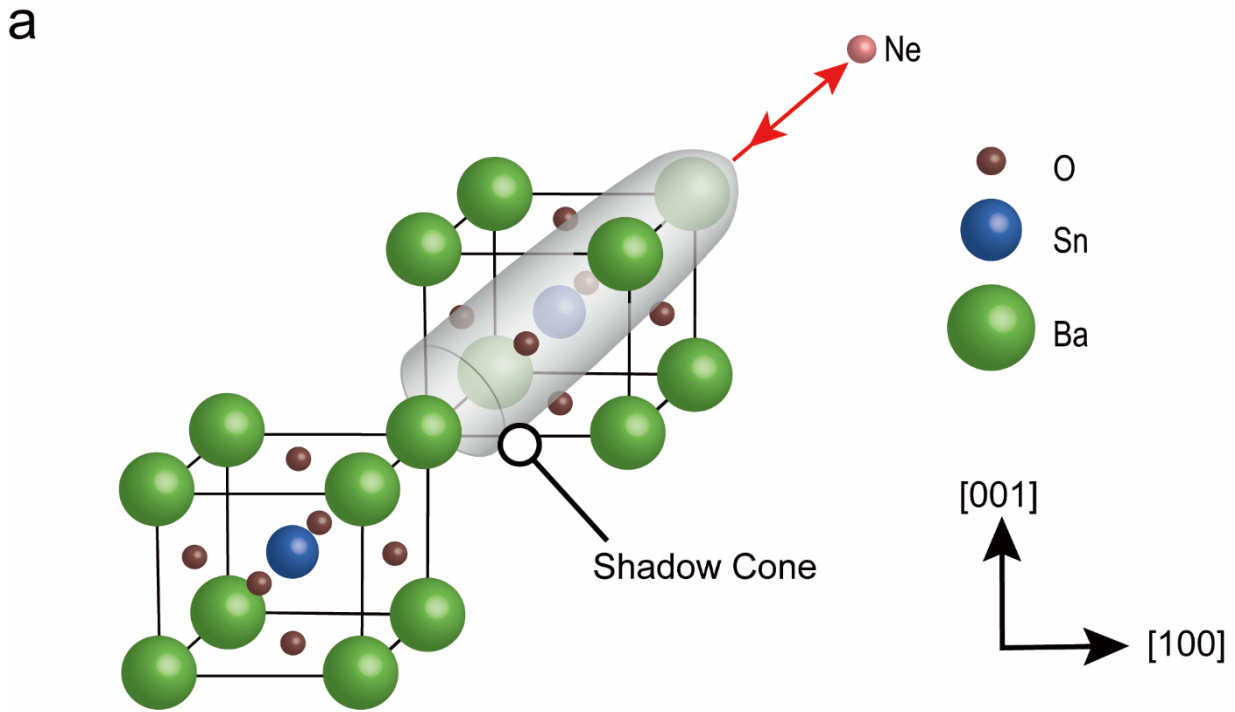

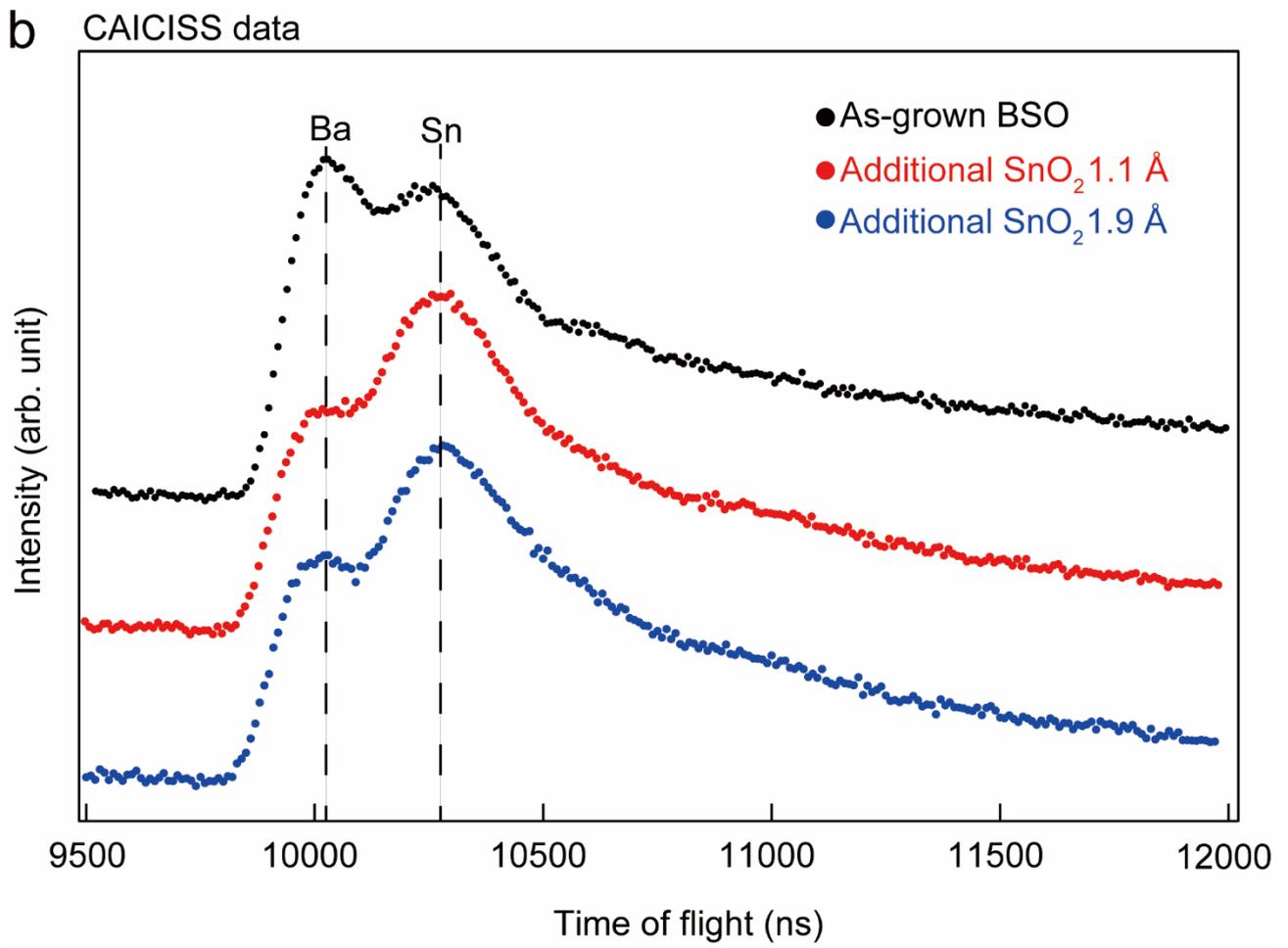

**Figure 4.** CAICISS spectrum of three different samples (As grown BSO / additional 1.1 Å $SnO_2$ deposition on BSO / additional 1.9 Å $SnO_2$ deposition on BSO). a) Measurements were performed at [111] direction with a Ne ion beam. Ne ions are backscattered by the top most atomic layer, and the inner ions are hidden by the shadowing cone. b) The CAICISS spectrum of the As-grown BSO shows that the intensity of both the Ba and Sn peaks are quite large. The Ba peak intensity is significantly decreased in the additional $SnO_2$ spectrum. The additional $SnO_2$ layer deposited BSO samples are $SnO_2$ dominant. However, there are not a significant differences between 1.1 Å and 1.9 Å of additional $SnO_2$.

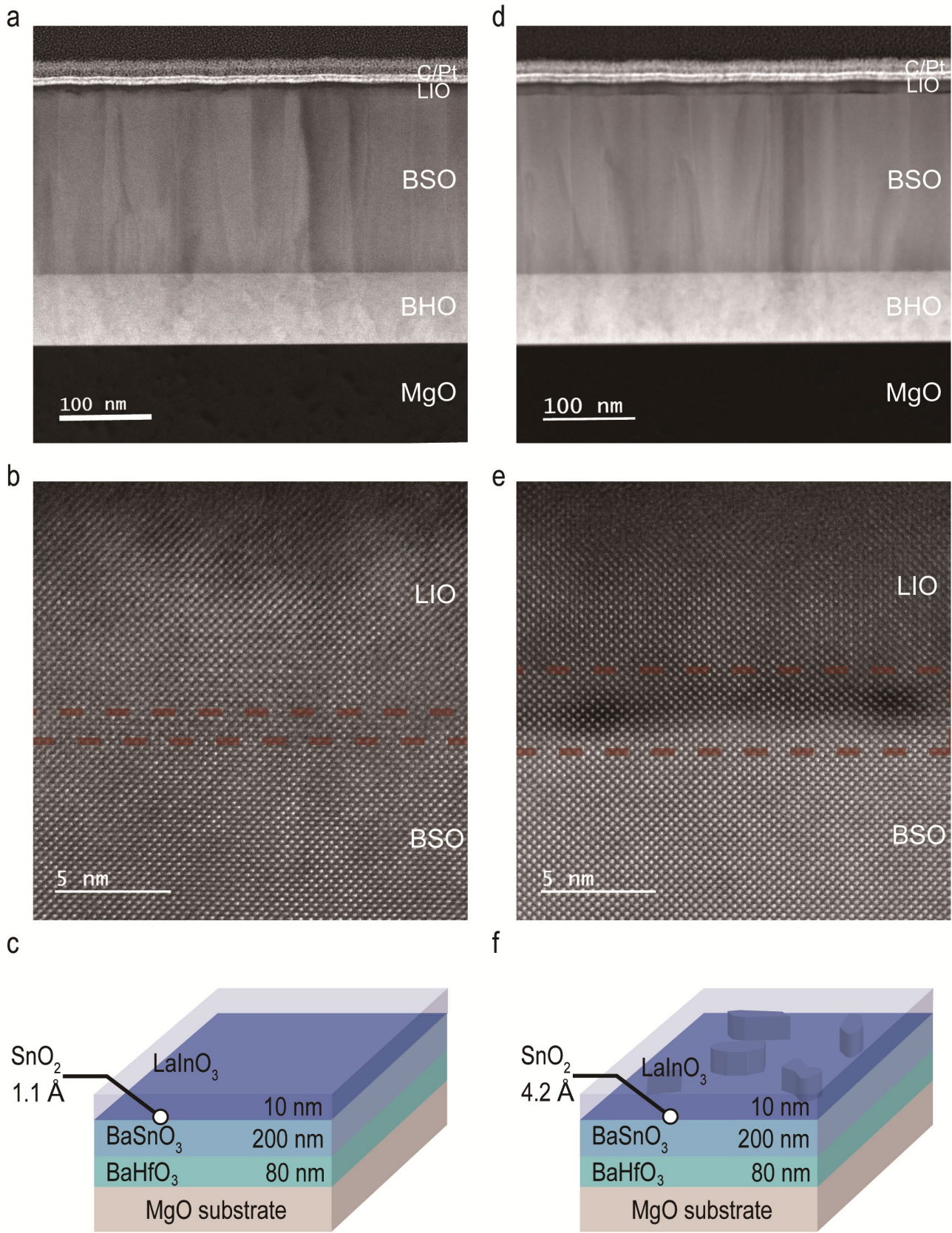

**Figure 5.** STEM images of BSO/LIO. (a, b) HAADF-STEM image of LIO on optimally SnO₂

terminated BSO (3 pulse dusting). c) Schematic diagram of LIO on SnO$_2$ terminated BSO film on an MgO substrate. (d, e) HAADF-STEM image of LIO with excessive SnO$_2$ (11 pulse dusting) on BSO. The red rectangular lines of (b) and (e) are guides for the interfacial region. f) Schematic diagram of LIO with excessive SnO$_2$ dusting on BSO film on an MgO substrate.